\documentclass[12pt]{article}

\pdfoutput=1

\usepackage{amssymb,graphicx}
\usepackage[intlimits]{amsmath}
\usepackage{bbm}
\usepackage[small]{subfigure}

\usepackage{MnSymbol}

\begin{document}

\thispagestyle{empty}

\begin{flushright}\footnotesize

\texttt{ICCUB-15-013}
\vspace{0.6cm}
\end{flushright}

\renewcommand{\thefootnote}{\fnsymbol{footnote}}
\setcounter{footnote}{0}

\def\caln{\mathcal{N}}

\begin{center}
{\Large\textbf{\mathversion{bold} Large $N_c$  
from Seiberg-Witten Curve\\ and Localization
}
\par}

\vspace{0.8cm}

\textrm{Jorge~G.~Russo}
\vspace{4mm}

\textit{ Instituci\'o Catalana de Recerca i Estudis Avan\c cats (ICREA), \\
Pg. Lluis Companys, 23, 08010 Barcelona, Spain}\\
\textit{   Department ECM, Institut de Ci\`encies del Cosmos,  \\
Universitat de Barcelona, Mart\'\i \ Franqu\`es, 1, 08028 Barcelona, Spain}
\vspace{0.2cm}
\texttt{jorge.russo@icrea.cat}

\vspace{3mm}


\par\vspace{0.4cm}

\textbf{Abstract} \vspace{3mm}

\begin{minipage}{13cm}
When ${\cal N}=2$ gauge theories are compactified on ${\mathbb S}^4$,
the large $N_c$ limit then selects a unique vacuum of the theory determined by saddle-point equations, which remains  determined even in the flat-theory limit.
We show that exactly the same equations can be reproduced  purely from Seiberg-Witten theory, describing   a vacuum where magnetically charged particles become massless, and corresponding to a specific
degenerating limit of the Seiberg-Witten spectral curve where $2N_c-2$ branch points join pairwise giving $a_{Dn}=0$, $n=1,...,N_c-1$.
We consider the specific case of  ${\cal N}=2$ $SU(N_c)$  SQCD coupled with $2N_f$ massive fundamental flavors. We  show that the theory exhibits 
a quantum phase transition where the critical point describes a particular  Argyres-Douglas point 
of the Riemann surface.

\end{minipage}

\end{center}

\vspace{0.5cm}


\newpage
\setcounter{page}{1}
\renewcommand{\thefootnote}{\arabic{footnote}}
\setcounter{footnote}{0}





\def\Xint#1{\mathchoice
   {\XXint\displaystyle\textstyle{#1}}%
   {\XXint\textstyle\scriptstyle{#1}}%
   {\XXint\scriptstyle\scriptscriptstyle{#1}}%
   {\XXint\scriptscriptstyle\scriptscriptstyle{#1}}%
   \!\int}
\def\XXint#1#2#3{{\setbox0=\hbox{$#1{#2#3}{\int}$}
     \vcenter{\hbox{$#2#3$}}\kern-.52\wd0}}
\def\ddashint{\Xint=}
\def\dashint{\Xint-}

\newcommand{\be}{\begin{equation}}\newcommand{\ee}{\end{equation}}
\newcommand{\bea}{\begin{eqnarray}} \newcommand{\eea}{\end{eqnarray}}
\def\sech{ {\rm sech}}
\def\p{\partial}
\def\pa{\partial}
\def\ov{\over }
\def\a{\alpha }
\def\g{\gamma}
\def\s{\sigma }
\def\td{\tilde }
\def\vp{\varphi}
\def\gd{\nu }
\def \ha {{1 \over 2}}

\def\KK{{\cal K}}
\def\stint{\strokedint}

\newcommand\cev[1]{\overleftarrow{#1}}


\section{Introduction}

Exact results in ${\cal N}=2$ theories have been obtained prominently  by three methods: the Seiberg-Witten approach \cite{Seiberg:1994rs,Seiberg:1994aj}, which computes the low energy effective action using  holomorphy, the instanton partition function in the $\Omega$-background
 \cite{Nekrasov:2002qd,Nekrasov:2003rj} 
and supersymmetric localization on ${\mathbb S}^4$ \cite{Pestun:2007rz} (or on other compact supersymmetric spaces) by which one can compute the free energy and the expectation value of 1/2 supersymmetric loop observables in terms of a matrix integral.
While these methods compute different quantities, under certain circumstances the results can be directly connected and compared.

Consider, for example, the case of ${\cal N}=2$ $SU(2)$ gauge theories on a four-sphere of radius $R$ \cite{Russo:2014nka}.
In the decompactification limit the localization partition function can be dominated by a saddle-point.
When this takes place, a vacuum is selected and the free energy $F=-\ln Z$ becomes proportional to  the holomorphic prepotential, ${\cal F}$, 
as $F=-R^2{\rm Re} (4\pi i {\cal F}) $,
including the instanton contributions. The prepotential is  evaluated at the singularity of the Seiberg-Witten curve where the dual magnetic variable
$a_D$ vanishes. Since $a_D=\partial {\cal F}/\partial a$, in turn this implies that the vacuum selected by minimizing the  free energy on  a large four-sphere
corresponds to the minimum of the (flat-theory) prepotential. Furthermore, in the case of $SU(2)$ SQCD one can show \cite{Russo:2014nka} that there is a quantum phase transition whose critical point exactly corresponds to the Argyres-Douglas \cite{Argyres:1995jj} superconformal point of the theory \cite{Argyres:1995xn}.

The quantum phase transition arising in the $SU(2)$ SQCD model is a low-rank analog of the large $N_c$ phase transitions discussed 
in \cite{Russo:2013qaa,Russo:2013kea} that arise in the decompactification limit for various massive ${\cal N}=2$ $SU(N_c)$ gauge theories.
The physical origin of these phase transitions can be  understood from the mass spectrum of the theory. In the background where the scalar field of the vector multiplet has expectation value
\begin{equation}\label{expPhi}
 \left\langle \Phi \right\rangle =\mathop{\mathrm{diag}}\left(a_1,\ldots ,a_{N_c}\right)\ ,
\end{equation}
the vector multiplets   have masses
 \begin{equation}\label{mv}
 m_{ij}^{\rm v}=\left|a_i-a_j\right|\ .
  \end{equation}
On the other hand, massive hypermultiplets in the adjoint and (anti) fundamental representation have masses
\begin{equation}
 \label{mh}
 m_{ij}^{\rm h, adj}=\left|a_i-a_j\pm M\right|,\qquad  m_{i}^{\rm h, f}=\left|a_i\pm M\right|\ .
\end{equation}
At large $N_c$, the $a_i$ are determined by minimizing the effective action. As a result, the hypermultiplet masses on this vacuum depend on the coupling and on the parameter $M$.
If the dynamics is such that for some finite, critical coupling some hypermultiplet masses vanish, the theory can undergo a phase transition. Specifically, this occurs as follows.
The effective action  consists of the classical piece and  one-loop corrections from integrating the different fields. In the large $N_c$ limit, instanton contributions can be neglected (in contrast with the
$N_c=2$ case, where the phase transitions are driven by instantons \cite{Russo:2014nka}).
The classical piece originates from the supersymmetric coupling of the scalar $\Phi $ to the curvature of ${\mathbb S}^4$,
\begin{equation}
 S_{\rm cl}=\frac{1}{4g_{\rm YM}^2 }\int_{{\mathbb S}^4}^{}d^4x\,\sqrt{g}\mathcal{R}\mathop{\mathrm{tr}}\Phi ^2\ .
\end{equation}
This gives
\be
\label{clasi}
 \frac{S_{\rm cl}[a]}{R^2} = \frac{8\pi ^2N}{\lambda }\sum_{i}^{}a_i^2\ .
\ee
(In asymptotically free theories, the coupling $\lambda $ is as usual traded by a dynamical scale $\Lambda $).We note that this term does not vanish in the large radius limit, despite the curvature goes
to zero; because of the volume factor it grows like $R^2$. On the other hand,
at large $R$, the one-loop contribution of a field of mass $m$ is proportional to  $m^2R^2\ln m^2R^2$,
where $1/R$ can be thought of as an infrared cutoff. 
The total effective action $S_{\rm eff}$ is then obtained by adding the classical term \eqref{clasi} to the one-loop contribution  summed over the mass spectrum \eqref{mv}, \eqref{mh} of the theory with appropriate coefficients.

At weak coupling $\lambda\ll 1$, eigenvalues are small, due to the quadratic potential in the classical term \eqref{clasi}.
 However, as $\lambda $ increases, the eigenvalues become larger, until some critical coupling at which they  hit the singularity where some of the hypermultiplets become massless,
that is, some   $m_{ij}^{\rm h, adj}$ or  $m_{ij}^{\rm h, f}$ vanish.
It turns out that the subcritical and supercritical solutions are different, leading to a discontinuity in the third (or higher) derivative of the free energy $F=S_{\rm eff}/R^2$ and
therefore a phase transition.
These phase transitions exhibit many striking features which have been further investigated in  \cite{Russo:2013sba,Chen:2014vka,Zarembo:2014ooa,Chen-Lin:2015dfa}.

\section{Super QCD in the Veneziano limit}



The Seiberg-Witten curve that describes ${\cal N}=2$ supersymmetric
$SU(N_c)$ gauge theory coupled to  fundamental hypermultiplets 
with arbitrary masses has been determined in \cite{Hanany:1995na,Argyres:1995wt}.
We are interested in the SQCD theory investigated  in \cite{Russo:2013kea,Russo:2013sba} using localization, which has $N_f$ fundamentals with  mass $M$ and $N_f$ antifundamentals with mass $-M$. In this case, the hyperelliptic curve is given by
\be
y^2= C(x)^2 -G(x)\ ,
\ee
\be
C(x)
= x^{N_c}+ \sum_{k=2}^{N_c} x^{N_c-k} s_k
\equiv  \prod_{i=1}^{N_c} (x-u_i)\ ,\qquad \sum_{i=1}^{N_c} u_i =0\ ,
\ee
\be
G(x)=\Lambda^{2N_c-2N_f} (x+M)^{N_f} (x-M)^{N_f}\ ,
\ee
where we consider $N_f< N_c$. Note that the total number of flavor multiplets, $2N_f$, is
even. In our notation, the superconformal case corresponds to $N_f=N_c$, $M=0$.

The meromorphic one-form is given by
\be
\lambda = x d\ln \frac{C-y}{C+y}\ .
\ee
The $a_n$, $a_{Dm}$ are periods of this differential form over a basis of homology one-cycles of the curve.
Consider the polynomial
\be
p(x)\equiv C(x)^2 -G(x)\ .
\ee
It has roots at $x_i,\ i=1,..., 2N_c$, which define the branch points of the curve.
Following \cite{Douglas:1995nw},  we  define $\alpha_m$, $m=1,...,N_c-1$, as the one-cycles that encircle $x_{2m}$ and $x_{2m+1}$,
and  $\gamma_i$, $i=1,...,N_c$, as  the one-cycles encircling $x_{2i-1}$ and $x_{2i}$. They satisfy
\be
\sum_i \gamma_i=0\ ,\qquad \langle \alpha_m,\gamma_j\rangle = \delta_{m,j}- \delta_{m,j+1}\ .
\ee
Monopoles are associated with cycles $\alpha_m$, and quarks, with cycles $\gamma_i$.
Let $\beta_n =\sum_{i\leq n} \gamma_i$ be  the cycles conjugate to $\alpha_m $.
Then
\be
a_n=\oint_{\beta_n} \lambda\ ,\qquad a_{Dm}=\oint_{\alpha_m} \lambda\ .
\ee
For $SU(N_c)$, in \cite{Douglas:1995nw}, the condition $a_{Dm}=0$ was used with the aim of studying the strong coupling
regime of pure ${\cal N}=2$ Super Yang-Mills theory.
For  $SU(2)$ gauge group, the condition $a_D=0$ 
defines a vacuum where the free energy and the prepotential are related by the formula  $F= -R^2{\rm Re}[4\pi i {\cal F}]$, including all instanton contributions \cite{Russo:2014nka}. 
 In this work we propose that $a_{Dm}=0$ defines  the unique vacuum selected at large $N_c$, once the ${\mathbb S}^4$ compactification has broken the vacuum
degeneracy (and $R\to \infty$ is taken afterwards). 
We note that this vacuum corresponds to the minimum of the prepotential:
\be
0=a_{Dm}=\frac{\partial {\cal F}}{\partial a_m}\ .
\label{adaa}
\ee
This is not a coincidence: in the $R\to\infty $ limit, 
solving the saddle-point equations indeed corresponds to minimizing the prepotential as long as there is a solution to the saddle-point equation
and the formula $F= -R^2{\rm Re}[4\pi i {\cal F}]$ holds.

To understand the implications of \eqref{adaa}, we first start with $N_f=0$, i.e. pure $SU(N_c)$ super Yang-Mills theory. The large $N_c$ limit of this theory 
was studied by Douglas and Shenker in \cite{Douglas:1995nw} and by Ferrari in \cite{Ferrari:2001mg}. Here we  give a different derivation.
Setting $N_f=0$ in the above formulas, we obtain
\be
p(x)\equiv y^2= C(x)^2 -\Lambda^{2N_c}\ ,\qquad C(x)=\prod_{i=1}^{N_c} (x-u_i)\ ,\ \ \sum_i u_i=0\ .
\ee
The condition $a_{Dm}=0$ requires that all $\alpha_m$ cycles shrink, $m=1,...,N_c-1$.
Namely, we must demand that $N_c-1$ roots of $p$ are double roots, so that 
 the curve takes the form
\be
y^2= (x-a)(x-b) \prod_{i=1}^{N_c-1} (x-c_i)^2\ .
\ee
This gives $N_c-1$ conditions, which completely fix the $u_i$ moduli parameters.
The general condition is that $p'$ shares the same roots $x=c_i$ as $p$. We have
\be
p' (x)= 
2\prod_{i=1}^{N_c} (x-u_i)^2 \sum_{i=1}^{N_c} \frac{1}{x-u_i}  =0\ .
\ee
Since none of the $u_i$ are roots, we find the condition
\be
\label{equm}
\sum_{i=1}^{N_c} \frac{1}{x-u_i}  =0\ .
\ee
which should hold for $x=c_i$, $i=1,...,N_c-1$.

In order to solve this equation at large $N_c$
we introduce as usual the (unit-normalized) eigenvalue density
\be
\rho(x) =\frac{1}{N} \sum_i \delta(x-u_i)\ .
\ee
Therefore, in the continuum, \eqref{equm} becomes
\be
\label{catorce}
\stint dy \ \frac{\rho(y) }{x-y} = 0\ .
\ee
This is indeed  the same  equation found from localization in \cite{Russo:2012ay} for pure  $SU(N_c)$ super Yang-Mills theory.
Typically, in the continuum  limit eigenvalues get distributed in cuts in the complex $x$-plane.
For the current theory, it turns out that there is a single cut on the real axes in some interval $(-\mu,\mu)$. We assume that the roots $c_i$ and the $u_i$ condense in the same interval $(-\mu,\mu)$ (modulo $1/N_c$ corrections).
This implies that  \eqref{catorce} must hold for any $x\in (-\mu,\mu)$.
Equation \eqref{catorce} then has a unique normalizable solution
\be
\label{densa}
\rho(x)= \frac{1}{\pi \sqrt{\mu^2-x^2}}\ .
\ee 
This is indeed the distribution first found by Douglas and Shenker \cite{Douglas:1995nw} by explicitly obtaining the curve with double roots
using Chebyshev polynomials, then computing the  periods $a_n$ and taking a scaling limit.

In order to determine $\mu $, we must also use the condition $p(x)=0$
for the roots. Taking the logarithm, we obtain
\be
\sum_{i=1}^{N_c} \ln (x-u_i)^2 = 2N_c \ln \Lambda \ .
\ee
In the continuum limit, this gives
\be
\label{quince}
\int_{-\mu}^\mu dy \ \rho(y) \ln (x-y)^2 = 2 \ln \Lambda\ .
\ee
Substituting the density \eqref{densa} into \eqref{quince} and computing the integral, we find $\mu=2\Lambda $.

\smallskip

Let us now consider the general case with $2N_f$ flavors.
We now get
\be
p' (x)= 
2\prod_{i=1}^{N_c} (x-u_i)^2 \sum_{i=1}^{N_c} \frac{1}{x-u_i}  - N_f \Lambda^{2N_c-2N_f} (x^2-M^2)^{N_f} \left( \frac{1}{x+M} + \frac{1}{x-M}\right)\ .
\ee
One particular solution of $p(x)=p'(x)=0$ is when some roots $u_i$ equal $\pm M$.
We will later see that this particular solution is associated with a supercritical regime.
Let us first find the generic solution.
Using $p(x)=0$, we can write $p'(x)$ in the following form
\be
\label{equn}
p' (x)= \Lambda^{2N_c-2N_f}  (x^2-M^2)^{N_f}\left(
2\sum_{i=1}^{N_c} \frac{1}{x-u_i}  - \frac{N_f}{x+M} - \frac{N_f}{x-M}\right)=0\ .
\ee
To take the large $N_c$ limit, we first introduce 
the Veneziano parameter
\begin{equation}
\zeta =\frac{N_f}{N_c}\, ,
\end{equation}
which remains fixed at $N_c\to\infty$ .
In the continuum,  \eqref{equn} gives
\be
\label{mfree}
2\stint_{-\mu }^{\mu } dy\,\,  \frac{\rho(y)}{x-y}
=   \frac{\zeta }{x+M} +\frac{\zeta }{x-M}\,.
\ee
Strikingly,  this is exactly  the second derivative of the saddle-point equation  found in \cite{Russo:2013kea,Russo:2013sba} from the localization partition function on ${\mathbb S}^4$ 
at large radius (see eq. (5.7) in \cite{Russo:2013kea}, or (4.7) in  \cite{Russo:2013sba}).
 The parameter $\mu $ is determined by demanding that the roots also solve $p(x)=0$:
\be
\prod_{i=1}^{N_c} (x-u_i)^2 = \Lambda^{2N_c-2N_f} (x+M)^{N_f}(x-M)^{N_f}\ .
\ee
Taking the logarithm and going to the continuum limit, we find
\be\label{aSQCD}
2\int_{-\mu }^{\mu } dy\,\rho(y)\ln \frac{\left(x-y\right)^2}{\Lambda ^2}
= \zeta \ln\frac{\left(x^2-M^2\right)^2}{\Lambda ^4}\, ,
\ee
which exactly reproduces eq. (4.6) in \cite{Russo:2013sba}, representing the first derivative of the large $N_c$ saddle-point equation
in the partition function on large ${\mathbb S}^4$.

It is interesting to see how these equations arise from the localization partition function given in \cite{Russo:2013kea,Russo:2013sba}.
Using the asymptotic formula for the Barnes $G$-function, we find that, at large radius,  the partition function
takes the form 
\be
Z^{\rm QCD}_{2N_f} =\int d^{N-1}a\ e^{-S(a)}\ ,
\ee
where 
\bea
\frac{S(a)}{R^2} 
&=& \sum_i \Big( -2(N_c-N_f) \big(\ln\Lambda R+\frac{3}{2}\big)\ 
 a_i^2 - \frac{N_f}{2} (a_i+M)^2 \ln (a_i+M)^2 R^2
\nonumber
\\
&-&  \frac{N_f}{2} (a_i-M)^2\ln (a_i-M)^2R^2\Big)+\frac{1}{2}\sum_{i,\ j} (a_i-a_j)^2\ln (a_i-a_j)^2R^2\ , \label{azio}
\eea
where the terms with logarithms represent the one-loop contribution and the first term
comes from the classical coupling to the curvature, as described earlier. As mentioned, instantons are negligible in the large $N_c $ limit.
Differentiating with respect to $a_i$, we  obtain the saddle-point equations
\bea
&& -4(1-\zeta) \big(\ln\Lambda R +1\big)\  a_i -\zeta  (a_i+M)\ln (a_i+M)^2R^2- \zeta  (a_i-M)\ln (a_i-M)^2R^2
\nonumber\\
&&+\frac{2}{N_c}\sum_{j\neq i} (a_i-a_j)\ln (a_i-a_j)^2R^2=0\ .
\label{artp}
\eea
By further differentiating with respect to $a_i $, one obtains an equation which in the continuum limit reduces to \eqref{aSQCD}.
Another differentiation then leads to \eqref{mfree} (see section 3).

The action \eqref{azio} is  proportional to the one-loop prepotential ${\cal F}$ \cite{D'Hoker:1996nv} of the theory, $S= -R^2{\rm Re}[4\pi i {\cal F}]$, ignoring the instanton part. Minimizing the action is therefore equivalent
to the condition $a_{Di}=0$, which minimizes the prepotential. However, there is a conceptual difference:
in the partition function one integrates over $a_i$; in the prepotential the $a_i$ label
different vacua. At large $N_c$, the full integration over $a_i$ is exactly determined by the saddle-point calculation, which selects the  particular vacuum that minimizes $S$.
In other words, the  large $N_c$ dynamics of the theory on a large ${\mathbb S}^4$ selects the vacuum described by the singular curve with $a_{Di}=0$.

The  theory depends only on two parameters, $\zeta $ and $\Lambda/M$, representing the coupling.
The terms on the RHS of \eqref{mfree} have poles at $x=\pm M$ which may or may not lie within the eigenvalue distribution, depending on the value of  $\Lambda/M$. The poles are associated with massless hypermultiplets which appear in the spectrum as soon as the eigenvalue distribution spreads over the singularities at $x=\pm M$.
Therefore the theory exhibits two different solutions, giving rise to two phases: the weak-coupling phase with $\mu <M$, in which all hypermultiplets are heavy, and the strong-coupling phase at $\mu >M$, where massless hypermultiplets appear in the spectrum. Let us briefly 
review these solutions \cite{Russo:2013kea}.

\subsubsection*{Weak-coupling phase ($\mu < M$):}

The poles at $x=\pm M$ sit outside the eigenvalue distribution. The solution to \eqref{mfree} is then 
given by
\begin{equation}
\label{wara}
 \rho (x)=\frac{1}{\pi \sqrt{\mu ^2-x^2}}\left(1-\zeta +\frac{\zeta M\sqrt{M^2-\mu ^2}}{M^2-x^2}
 \right).
\end{equation}
Substituting this solution into \eqref{aSQCD},  we find a transcendental equation for $\mu $.  The resulting $\mu $ can be expressed in a parametric form:
\begin{eqnarray}\label{muuu}
\mu &=&M\sqrt{1-u^2},
\\
 \left(\frac{2\Lambda }{M}\right)^{2-2\zeta }&=&\left(1+u\right)^{1-2\zeta }\left(1-u\right).
 \label{udefinition}
\end{eqnarray}
As $\Lambda/M$ is increased from zero, $\mu $ eventually reaches $M$. This occurs at $2\Lambda = M$.
Note that for $\zeta=1/2$, equations simplify, giving
\be
\mu =2\sqrt{\Lambda (M-\Lambda)}\ .
\ee

\subsubsection*{Strong-coupling phase ($\mu >M$):}

The poles now sit within the eigenvalue distribution. The  eigenvalue density is then given by
\be
\label{noj}
\rho (x) = \frac{1-\zeta }{\pi \sqrt{\mu^2-x^2}} + \frac{\zeta }{2}\,\delta(x+M)+\frac{\zeta }{2}\,\delta(x-M).
\ee
  Substituting the solution (\ref{noj}) into \eqref{aSQCD}, we now obtain:
\begin{equation}\label{mu=2Lambda}
 \mu =2\Lambda .
\end{equation}
Thus the strong coupling phase occurs at $2\Lambda > M$.

\bigskip

The distribution \eqref{noj} has   $N_f/2$ double roots located at $x=\pm M$ and $N_c-N_f$ double roots distributed in the interval 
$(-\mu,\mu)$. Thus in this phase the Seiberg-Witten curve has the form
\be
\label{carta}
y^2 = (x+M)^{N_f}  (x-M)^{N_f} \prod_{i=1}^{N_c-N_f} (x-c_i)^2\ .
\ee
This implies that, in the supercritical regime, many cycles have collapsed to
give rise to a particular Argyres-Douglas point on the Coulomb branch. At this point, some mutually non-local states
become massless. In the   curve \eqref{carta}, not only all $\alpha_m$ cycles are collapsed but also a number $N_f$ of $\beta_n$ cycles are collapsed as well.
For $SU(N_c)$ SQCD,  similar critical points were investigated in \cite{Eguchi:1996vu},   describing a theory with an infrared-free $SU(2)$ gauge multiplet  coupled to two different superconformal theories
\cite{Gaiotto:2010jf} (see also \cite{Giacomelli:2012ea,Giacomelli:2013tia}).
It would of great interest to understand the IR structure of the theory described by the particular singular curve \eqref{carta}.

The special case $N_f=N_c$ (i.e. $\zeta=1$) corresponds to a massive deformation of the ${\cal N}=2$ superconformal theory. In this case, no phase transition occurs, the theory  always stays in the phase with $\mu<M$ \cite{Russo:2013kea}
\begin{equation}
\label{wara}
 \rho (x)= M\sqrt{M^2-\mu ^2}\ \frac{1}{\pi \sqrt{\mu ^2-x^2}\ (M^2-x^2)}\ ,\qquad \mu =\frac{M}{\cosh \frac{4\pi^2}{\lambda }} .
\end{equation}

\section{Solving the discrete saddle-point equations}

 The saddle-point equation \eqref{artp} determines the eigenvalues of the scalar field $\Phi $ that minimize the action. In this subsection we will
show that it admits a solution where the eigenvalues are the zeros  of an associated Legendre polynomial.

Differentiating \eqref{artp} with respect to $a_i$, we find
\be
 4(1-\zeta) \ln\Lambda R +\zeta  \ln (a_i+M)^2R^2+ \zeta \ln (a_i-M)^2R^2
-\frac{2}{N_c}\sum_{j\neq i} \ln (a_i-a_j)^2R^2=0\ .
\label{artos}
\ee
Another differentiation with respect to $a_i$ gives
\be
\label{haya}
\frac{1}{N_c}\sum_{k\neq i} \frac{2}{a_i-a_k} = \frac{\zeta}{a_i-M} +\frac{\zeta}{a_i+M}\ .
\ee
Discrete saddle-point equations of this type can be exactly solved for any  finite $N_c$   by a method introduced by Stiltjes \cite{stieltjes}. The method was more recently applied to the context of ${\cal N}=4$ theory in \cite{Lubcke:2004dg}. 
Following \cite{Lubcke:2004dg}, we define 
\be
Q(x)\equiv \prod_{i=1}^{N_c} (x-a_i)\ .
\ee
Then
\be
Q'(a_k) = \prod_{i\neq k} (a_k -a_i) \ ,
\ \ \ Q^{\prime\prime}(a_k)= \prod_{i\neq k} (a_k-a_i) \sum_{j\neq k}\frac{2}{a_k-a_j}\ .
\ee
Using \eqref{haya}, we get
\be
P(a_k)\equiv (a_k^2-M^2)Q^{\prime\prime} (a_k) -2N_f a_k Q'(a_k)  =0\ .
\ee
Since $P(x)$ is a polynomial of order $N_c$ with the same roots as $Q(x)$, then $P$ and $Q$  can only differ by an overall  coefficient. This is found by comparing the $x^{N_c}$ term:
\be
P(x) = N_c(N_c-1-2N_f)   x^{N_c}+...
\ee
Hence we obtain the differential equation
\be
(x^2-M^2) Q^{\prime\prime} - 2N_f x Q' - N_c(N_c-1-2N_f)  Q = 0\ .
\ee
The solution is expressed in terms of associated Legendre polynomials. The polynomial solution is
\be
\label{aswa}
Q(x)=b\ \left(1-\frac{x^2}{M^2}\right)^{\frac{N_f+1}{2}} P_{\ell }^{N_f+1}(x/M)\ ,\qquad \ell \equiv N_c-N_f-1\ ,
\ee
or
\be
\label{brw}
Q(x) =c\ \left( 1-\frac{x^2}{M^2} \right) ^{N_f+1}  \frac{d^{N_c}}{dx^{N_c}}\left( \frac{x^2}{M^2}-1 \right) ^{N_c-N_f-1}\ ,
\ee
where $b,\ c$ are  irrelevant numerical coefficients. This last formula shows
that $Q$ is a polynomial in both cases, $N_f$ even and $N_f$ odd.
It should be noted that the representation  \eqref{brw} only holds for $\ell\ge N_f+1$, i.e.
$N_f\leq N_c/2-1$. For $N_f>N_c/2-1$, $Q$ is given in terms of an associated Legendre function and it also simplifies to a polynomial.

By construction, the eigenvalues are the $N_c$ roots of $Q$.
For $N_f\leq N_c/2-1$, $N_f+1$ zeros are located at $x=\pm M$.
The remaining $N_c-2N_f-2$  zeros can be identified with the remaining roots of the associated Legendre polynomial. Thus this solution has $N_f+1$ eigenvalues piling up at $\pm M$ and  the rest lying in the interval $(-M,M)$. Therefore it describes the critical case. In this  particular  solution,  $\Lambda/M$ is determined from \eqref{artos}
in terms of $N_c,\ N_f$, generalizing the critical, large $N_c$ relation $2\Lambda=M$ to finite $N_c,\  N_f$. 
It would be very interesting to find the discrete solution for  arbitrary coupling $\Lambda/M$.

\section{Conclusion}

To summarize, in this note we have  described  how to extract  the large $N_c$ dynamics from
the Seiberg-Witten curve, when a four-sphere is used as an infrared regulator.
This adds a new physical interpretation on the properties of the Seiberg-Witten curve: 
there is a special degenerating limit defined by the condition  $a_{Di}=0$, $i=1,...,N_c-1$, where $N_c-1$ branch points join pairwise, making all  $\alpha_i $ cycles shrink to zero.
In this limit, the periods $a_i$  describe the unique vacuum selected by the ${\mathbb S}^4$ compactification in the large radius limit.
At large $N_c$, the condition $a_{Di}=0$ leads to integral equations which are exactly the same as the
integral equations that determine the saddle-point of the localization path integral. This is somewhat surprising, since the origin of these equations is very different; in one case, they arise by going to a specific degenerating limit of the Seiberg-Witten curve; in the other case, they arise by minimizing the effective action on ${\mathbb S}^4$, {\it including} the supersymmetric  coupling of the scalar field to the curvature.
The structure of the vacuum is then determined by the integral equations. 
Different solutions may appear at different intervals of the coupling, describing different  phases of the theory. At criticality, at least a pair of conjugate homology cycles shrink simultaneously. These represent Argyres-Douglas points of the curve, where mutually non-local states become massless.

\subsection*{Acknowledgements}
 
We would like to thank K. Zarembo for useful comments and for bringing our attention
to references \cite{stieltjes,Lubcke:2004dg}. We also thank Y. Tachikawa for useful remarks.
We acknowledge financial support from projects  FPA2013-46570  and
 2014-SGR-104.



\end{document}